# Cooperative Spectrum Sensing with Enhanced Energy Detection under Gaussian Noise Uncertainty in Cognitive Radios

*HUANG He*, et.al

Beijing University of posts and telecommunications, Beijing, 100876, China

**Abstract:** This paper presents optimization issues of energy detection (ED) thresholds in cooperative spectrum sensing (CSS) with regard to general Gaussian noise. Enhanced ED thresholds are proposed to overcome sensitivity of multiple noise uncertainty. Two-steps decision pattern and convex samples thresholds have been put forward under Gaussian noise uncertainty. Through deriving the probability of detection ($P_d$) and the probability of false alarm ($P_f$) for independent and identical distribution (*i.i.d.*) SUs, we obtain lower total error rate ($Q_e$) with proposed ED thresholds under low signal-to-noise-ratio (SNR) condition. Furthermore, simulation results show that proposed schemes outperform most other noise uncertainty plans.

## 0 Introduction

Nowadays, (CRs) popular non-coherent spectrum sensing (SS) algorithm---energy detection (ED) has an advantage in low computational complexity and it does not need any knowledge of signal[1-4] in cognitive radios. However ED algorithm could not avoid noise uncertainty due to fixed ED threshold. Furthermore, it is very difficult to implement ED when the signal-to-noise (SNR) is below a fixed value[3]. In Ref. [5] dynamic threshold is put forward to make detection probability higher without increasing computation cost, but small change of noise will lead to the inaccurate performance. Double-thresholds algorithm has been suggested to predict the state of primary user (PU)[3], but it only considers noise variation between maximum value and average value, and this algorithm could not obtain lower total error rate ($Q_e$). Aiming at optimizing the probability of detection ($P_d$) with noise uncertainty, an adaptive detective model is proposed, but the probability of false alarm ($P_f$) is close to 1 as $P_d$ increase to 1[6]. To sum up, it is a critical problem to make the $Q_e$ lower enough under noise uncertainty in low SNR environment.

According to the problems above, in this letter we put forward enhanced ED thresholds to mitigate influence of noises variation in cooperative spectrum sensing (CSS), and optimize ED thresholds under Gaussian noises which are represented by the complex vector matrix. We improve $P_d$ and ensure $Q_e$ low enough. At last, simulation results show that the proposed plans have better performance than other noise uncertainty algorithms.

## 1 System model

The binary hypothesis test for detection of an unknown stochastic received signal ($H_0$: signal is absence; $H_1$: signal is present) is formulated as Ref. [6],

$$H_0 : y(t) = n(t)$$
$$H_1 : y(t) = h \cdot s(t) + n(t) \tag{1}$$

where $y(t)$ is the received signal at the secondary receiver, $h$ denotes the wireless channel gain, $s(t)$ is defined as required detective primary signal with variance $\sigma_s^2$, $n(t)$ is circularly symmetrical complex additive white Gaussian noise (AWGN) of variance $\sigma_c^2$. The ED model could be expressed as,

$$Y = \frac{1}{k}\sum_{i=1}^{k}\frac{y^2(i)}{\sigma_c^2} \qquad (2)$$

where $k$ is the number of samples used for computation, and $Y$ is the test statistic compared with ED threshold $\gamma$. If $Y \geq \gamma$, the decision is $H_1$, otherwise $H_0$. The probability density function $(p.d.f.)$[7] of $y(t)$ can be given by Eq.(3),

$$f_Y(y) = \begin{cases} \dfrac{1}{2^{\frac{k}{2}}\Gamma(\frac{k}{2})} y^{\frac{k}{2}-1} e^{-\frac{y}{2}}; & H_0 \\ \dfrac{1}{2}\left(\dfrac{y}{SNR}\right)^{\frac{k}{4}-\frac{1}{2}} e^{-\frac{y}{2}-\frac{SNR}{2}} I_{\frac{k}{2}-1}(\sqrt{y \cdot SNR}); & H_1 \end{cases} \qquad (3)$$

where SNR is the signal-to-noise ratio, $I_{u-1}$ is the first kind modified Bessel function with the order $u$-1, $\Gamma(.)$ is the Gamma function and $\Gamma(.,.)$ is the incomplete gamma function. Accordingly, we could compute $P_d$ and $P_f$[7] as,

$$P_d = P_r(y > \gamma \mid H_1) = Q_u(\sqrt{2 \cdot SNR}, \sqrt{\gamma}) \qquad (4)$$

$$P_f = P_r(y > \gamma \mid H_0) = \frac{\Gamma(u, \frac{\gamma}{2})}{\Gamma(u)} \qquad (5)$$

where $u$ is the time bandwidth product and $Q_u(a,b)$ is the $u$-th order generalized Marcum Q-function,

In CSS the $i$-th second user (SU) makes a binary decision $D_i$ to fusion center (FC) on the basis of logical rules ("1" represent $H_1$, "0" represent $H_0$), where all 1-bits have been added up together. In Eq.(6) "$n$-out-of-$K$" optimal voting rule is represented in CSS, $D_i$ stands for binary decision of $H_1$ and $H_0$, $K$ is the number of SUs and $v$ is the integer, it is shown that $v=1$ for OR rule and $v=K$ for AND rule[8].

$$\sum_{i=1}^{K} D_i \begin{cases} \geq v, & H_1 \\ < v, & H_0 \end{cases} \qquad (6)$$

## 2 Enhanced energy detector schemes $\gamma$, $\gamma$` and $\gamma$``

We consider there are $m(1 \leq q \leq m)$ SUs, a common PU and one FC in CRs networks as in Fig.1. Each SU is assumed independently. Binary hypothesis test is expressed as follows,

$$\begin{aligned} H_0 &: y_{q\tau}(t) = n_{q\tau}(t) \\ H_1 &: y_{q\tau}(t) = h \cdot s_\tau(t) + n_{q\tau}(t) \end{aligned} \qquad (7)$$

where $\tau$ is signal component index ($\tau=1,2,...,k$) at the $q$-th SU and $t$ is time instant, $n_{q\tau}(t)$ is the $\tau$-th noise component in the $q$-th SU noise of variance $\sigma_c^2$. $s_\tau(t)$ is the received circularly symmetrical complex AWGN with variance $\sigma_\tau^2$. For each SU with $k$ signal component indexes contains independent optimized energy detector.

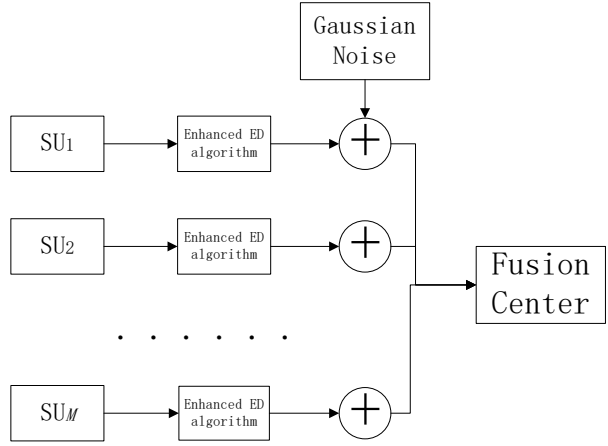

Fig.1 The system model of CSS with enhanced ED under Gaussian noise

The system model with enhanced ED in CSS under Gaussian noise is presented in Fig.1. The enhanced ED scheme $\gamma$ can be shown as ($1 \leq d \leq m$),

$$\gamma_{\min} = \min \sum_\tau \frac{y_{m\tau}^2}{\sum_\tau \frac{\sigma_{m\tau}^4}{\sigma_{d\tau}^2}}, \gamma_{\max} = \max \sum_\tau \frac{y_{m\tau}^2}{\sum_\tau \frac{\sigma_{m\tau}^4}{\sigma_{d\tau}^2}} \tag{8}$$

It can be inferred from Eq. (8) if the actual calculated value is greater than the least upper bound $\gamma_{max}$, we will decide $H_1$ and determine $H_0$ for computation result is less than the greatest lower bound $\gamma_{min}$.

Distribution threshold decision is not to be taken into account if the computation value is belong to [$\gamma_{min}$, $\gamma_{max}$], then we consider re-determining in [$\gamma_{min}$, $\gamma_{max}$], in order to reduce computational complexity, the enhanced ED scheme $\gamma`$ can be expressed as Eq. (9),

$$\gamma' = \sum_\tau \frac{y_{m\tau}^2}{\sum_\tau \frac{\sigma_{m\tau}^4}{E(\sigma_{m\tau}^2)}} \tag{9}$$

where $E(.)$ denotes the expectation.

On the other hand, to further balance noises variance effect for multiple sum of $\sigma_i^2$, weighted convex samples could be constructed to estimate the overall converges that is apparently upper convex function in Eq. (10). The series estimation with VOTING PROGRESS[8] is assumed to minimize the detection threshold with multiple Gaussian noises.

$$\gamma'' = \sum_\tau \frac{y_{m\tau}^2}{\sum_\tau \frac{\sigma_{m\tau}^4}{\min \frac{\sum_i^\tau \mu_{\tau-i}^g E(\sigma_{m\tau}^2)}{\sum_i^\tau \mu_{\tau-i}^g}}} \tag{10}$$

where $\mu_{\tau-i}$ denotes the expectation of Gaussian noise in Part 3 and $g$ denotes the positive integer. Furthermore, when $\mu_{\tau-i}=1$, the convex sample threshold in Eq. (10) could be simplified to Eq. (9).

## 3  General complex Gaussian noise simplification

In Fig.1 in order to simplify general complex random Gaussian noise, the signal can be represented as $x_i \sim N(\mu_i, \sigma_i^2)$, we define statistical directed matrix vector as $X=[x_1,x_2,...,x_n]$ and the transpose conjugate vector as $X^H=[x_1,x_2,...,x_n]^H$, the signal covariance matrix can be given by[6,9],

$$N = X \cdot X^H \tag{11}$$

then we get the expectation of noise,

$$E(\sigma_{i\mu}^2) = E(x_{i\mu} \cdot x_{i\mu}^H) - E(x_{i\mu})E(x_{i\mu}^H) \tag{12}$$

then we have,

$$E(\sigma_{\tau k}^2) = E(\sum_{i=1}^{n} \frac{(X-U)\cdot(X-U)^H}{\sigma_i^2}) = \sum_{i=1}^{n} \frac{\sum_\tau \sum_k E(x_{\tau k} x_{\tau k}^H - x_{\tau k}\mu_{\tau k}^H - \mu_{\tau k} x_{\tau k}^H + \mu_{\tau k}\mu_{\tau k}^H)}{\sigma_i^2} = E(\sum_{i=1}^{n} \frac{X \cdot X^H + U \cdot U^H}{\sigma_i^2}) \tag{13}$$

averaging noises as,

$$E(\sigma_\tau^2) = \frac{1}{n} E(\sum_{i=1}^{n} \frac{X \cdot X^H + U \cdot U^H}{\sigma_i^2}) \tag{14}$$

Furthermore, we consider if the sampling signals are collected in the intermediate interval of the Gaussian signal, we let confidence level is 1-$\alpha$, then we get,

$$\{\sigma^2\} \subset \left[\bar{\mu} - \kappa_{\frac{\alpha}{2}}\sigma, \bar{\mu} + \kappa_{\frac{\alpha}{2}}\sigma\right], \kappa_{\frac{\alpha}{2}} = \frac{u_{\frac{\alpha}{2}}}{\sqrt{n}}, k = \left[\sqrt{2\kappa_{\frac{\alpha}{2}}\sigma}\right] \tag{15}$$

where $\bar{\mu} = \sum_{i=1}^{n} u_i / n$, $\sigma^2$ follows stationary Gaussian uniform distribution. When 1-$\alpha$=0.99, then $\kappa_{\alpha/2}$=2.58. If we define 1-$\alpha$=0.8 to arrange $\sigma_{\tau k}^2$ for optimizing threshold, then,

$$\{\sigma^2\} \subset \left[(\bar{\mu} - \kappa_{\frac{0.01}{2}}\sigma)\frac{\kappa_{\frac{0.2}{2}}}{\kappa_{\frac{0.01}{2}}}, (\bar{\mu} + \kappa_{\frac{0.01}{2}}\sigma)\frac{\kappa_{\frac{0.2}{2}}}{\kappa_{\frac{0.01}{2}}}\right] \tag{16}$$

Noise simplification can be estimated via substituting Eq. (14) and Eq. (16) and it can be used for providing to appropriate detective threshold of multi-noises power.

On the other hand, for conventional ED plan, it requires about $k(\tau-1)$ multiplications. The calculation of two-steps scheme $\gamma$ is $\alpha_1 k(\tau-1)^2 + \alpha_2(k(\tau-1)^2 + o(k\tau))$, where $\alpha_1$ and $\alpha_2$ are $P_d$-related weight constant for twice decision. In addition, the computational complexity of limitation optimization solution $\gamma``$ is $k(\tau-1)^2 + o(k\tau^2)$, it could cut down $Q_e$, which is shown in section 5.

## 4 Optimization of CSS

In Fig.1 with CSS we get conventional density function (c.d.f.) of false alarm probability $P_f$ and missed detection probability $P_m$ based on binary hypothesis testing as Ref. [10],

$$p_{f,\tau} = p_r(H_1 | H_0) = 1 - \int_0^\gamma f_{y|H_0}(y)dy \tag{17}$$

$$p_{m,\tau} = p_r(H_0 | H_1) = 1 - \int_0^\gamma f_{y|H_1}(y)dy \tag{18}$$

From Ref. [10, Eq. (6)] we derive p.d.f. of $P_d$ and $P_f$ over independent and identically distributed (i.i.d.) SUs towards Gaussian noise uncertainty with Eqs. (8)- (10), Eq. (17) and Eq. (18), the p.d.f. of $P_d$ and $P_f$ can be shown as,

$$P_r(\frac{y_i^2}{(w^{\frac{1}{2}})^p} \leq y) = \frac{2}{p} y^{\frac{2}{p}-1} f_{\frac{y_i^2}{(w^{\frac{1}{2}})^p}}(y^{\frac{2}{p}}) \tag{19}$$

with mathematical induction for Eq. (9), when $p=2$,

$$P_r(\frac{y_i^2}{w} \leq y) = f_{\frac{y_i^2}{w}}(y) \tag{20}$$

$$f_{w|H_0}(y) = \frac{\partial P_r(\frac{y_i^2}{w} \leq y)}{\partial y} = \frac{\exp(-\frac{y}{w})}{w} \tag{21}$$

Similarly, we have,

$$f_{w|H_1}(y) = \frac{\exp(-\frac{y}{w(1+\overline{SNR})})}{w(1+\overline{SNR})} \tag{22}$$

$\overline{SNR}$ denotes the average signal-to-noise ratio.

The $P_m$, $Q_f$ and $Q_m$ of $k$ SUs in CSS can be expressed as Ref. [8],

$$p_m = 1 - p_d \tag{23}$$

$$Q_f = \sum_{l=n}^{k} \binom{k}{l} p_f^l (1-p_f)^{k-l} \tag{24}$$

$$Q_m = 1 - \sum_{l=n}^{k} \binom{k}{l} p_d^l (1-p_d)^{k-l} \tag{25}$$

where $l$, $n$ and $k$ are positive integers.

Keeping $Q_f$ and $Q_m$ as low as possible, we let $Q_e$ with weighted coefficients ($p_r(H_0)$ and $p_r(H_1)$) of the probability of binary hypothesis test as,

$$Q_e = p_r(H_0) \cdot Q_f + p_r(H_1) \cdot Q_m \tag{26}$$

in Eq. (26) we define $p_r(H_0)=\alpha$ and obviously $p_r(H_0)+p_r(H_1)=1$, then we get,

$$Q_e = \sum_{\sigma_i^2} P_r(H_0) \cdot P_r(\sigma_i^2 | H_0) Q_f + P_r(H_1) \cdot P_r(\sigma_i^2 | H_1) Q_m \tag{27}$$

To derive Eq. (27) with reflection on multiple noises function for $\sigma_i^2(Q_f)$ and $\sigma_i^2(Q_m)$ using Beyesian Approach respectively, we get,

$$Q_e' = \sum_{\sigma_i^2(\bullet), P_r(H_0, \sigma_i^2(Q_f)) \cdot Q_f = P_r(H_1, \sigma_i^2(Q_m)) \cdot Q_m} P_r(H_0) \cdot P_r(\sigma_i^2(Q_f) | H_0) \cdot Q_f + P_r(H_1) \cdot P_r(\sigma_i^2(Q_m) | H_1) \cdot Q_m \tag{28}$$

Substituting Eqs. (17)-(27) into Eq. (28) and taking integral derivative to gain the most optimized value of $N^*$ as differentiating Eq. (27), aiming at $N^*$ ($N^* = k-n$) in Eq. (24) and Eq. (25), we have,

$$N^* = \arg\min_N (\alpha \frac{\partial Q_f}{\partial N} + (1-\alpha) \frac{\partial Q_m}{\partial N}) = \arg\min_N (\alpha \frac{\partial Q_f}{\partial N} - (1-\alpha) \frac{\partial Q_d}{\partial N}) \tag{29}$$

For $Q_d(SNR,\gamma,N,\sigma_i^2)$ and $Q_f(u,\gamma,N,\sigma_i^2)$ have different values for $SNR$ and $u$, then we get,

$$N^* = \left\lceil \arg \frac{\min \alpha \cdot \frac{\partial Q_f}{\partial N}}{\max(1-\alpha) \cdot \frac{\partial Q_d}{\partial N}} \right\rceil = \left\lceil \arg \min_N \frac{\alpha \cdot \frac{\partial Q_f}{\partial N}}{(1-\alpha) \cdot \frac{\partial Q_d}{\partial N}} \right\rceil \quad (30)$$

where $\lceil \cdot \rceil$ expresses the ceiling function.

## 5  Simulation results

A comparison between the performances of proposed enhanced ED schemes and simulation of conventional ED models applied in multi-hop CRs networks and imperfect channel[10] (because noise uncertainty and general Gaussian noise can be valid for conditional ED threshold) are described in Fig.2 and Fig.3. Some other CSS noise uncertainty plans[11-13] also have been considered to be compared with given schemes for $P_m$ and $Q_e$ under low SNR condition in Eq. (27). OR-logic fusion rules have been adopted in the FC for CSS.

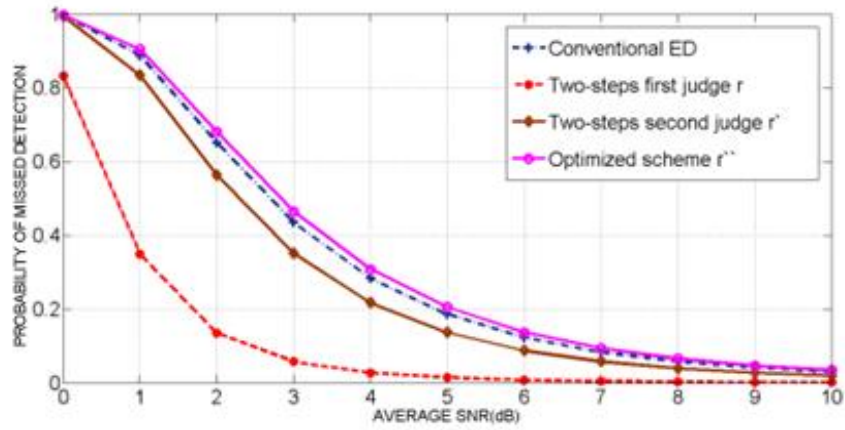

Fig.2 Probability of missed detection versus average SNR. $\gamma=30$, $N^*=5$(from Fig.2 in Ref. [8] the value selection of $\gamma$ and $N^*$ for ED programs could obtain the optimal $Q_e$), $w=E(\sigma_{m\tau}^2)$.(conventional ED in imperfect channel with error probability $q=0.001$)

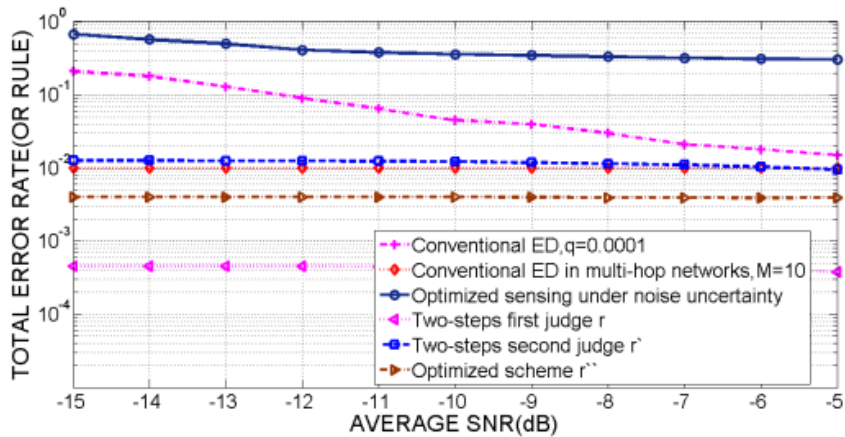

Fig.3  Total error rate (OR rule) of CSS versus average SNR for contrast between proposed plans and noise uncertainty algorithms. $\gamma=30$, $N^*=5$, $w=E(\sigma_{m\tau}^2)$.

Figure 2 and Figure 3 indicate that though scheme $\gamma$ have the best performance for $P_m$ as SNR increases and very low $Q_e$ at low SNR level, it requires secondary decision with scheme $\gamma`$. And convex optimization solution $\gamma``$ could once achieve rapid spectrum holes detection with no requirement for re-decision towards noise variable uncertainty. Furthermore, it is assumed that three optimized strategies have noticeable superiority in $Q_e$ than other noise uncertainty schemes.

On the other hand, we observe from Fig.4 when optimal cooperative CRs users increase and SNR=-10 dB, it is available to $Q_e$ become higher as the number of CSS increase, because the increased rate of $P_f$ is changing faster than the reduced rate of $P_m$. More cooperative users could improve $P_d$, but $P_f$ vary more quickly than $P_d$ in low SNR environment, which make $Q_e$ continue to become greater. Although scheme $\gamma$ has the lowest $Q_e$ as well as possible, it require re-determining with method $\gamma^{`}$ when computational value at the receiver is in the reconviction area. Besides, scheme $\gamma^{``}$ has nice performance and it can be conducted only one time. In addition, it can be observed that proposed algorithms are much less than the specified value 0.1 for the IEEE 802.22 cognitive wireless regional area network (WRAN) standard. Hence it is meaningful to employ suitable optimized energy detectors for signal detection under noise uncertainty in CSS.

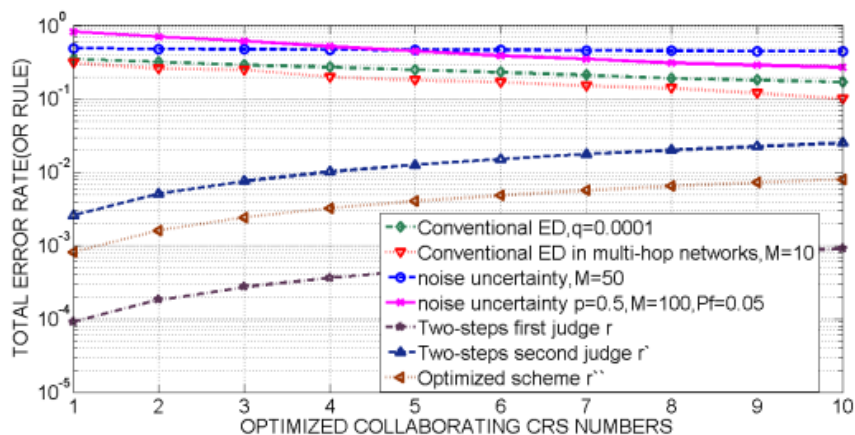

Fig.4  Total error rate (OR rule) versus optimal collaborating CRs number. SNR=-10dB, $\gamma$=30, $w=E(\sigma_{m\tau}^2)$

## 6  Conclusion

In this paper, we analyze the enhanced ED thresholds under general Gaussian noise in CSS. It is referred that we are able to effectively estimate Gaussian noises uncertainty with enhanced ED in low SNR condition. The proposed algorithms have moderate complexity and can be used for detecting the signal of SU more accurately and quickly in CSS.